\documentclass[onecolumn,floatfix,preprintnumbers,eqsecnum,letterpaper,superscriptaddress,nofootinbib]{revtex4}

\usepackage{graphicx}
\usepackage{microtype}
\usepackage{amsmath}
\usepackage{amssymb}
\usepackage{subfigure}
\usepackage{hyperref}
\usepackage{url}
\usepackage{xcolor}
\usepackage{color}
\usepackage{mathrsfs}
\usepackage{calrsfs}
\usepackage{amsfonts}
\usepackage{eufrak}
\usepackage{tabularx}
\usepackage{multirow}
\usepackage{eucal}
\usepackage{latexsym}
\usepackage{ragged2e}
\usepackage{epsfig}
\usepackage{textcomp}
\usepackage{caption}
\usepackage{subfigure}
\usepackage{marvosym}
\usepackage{ifsym}
\usepackage{lipsum}

\DeclareCaptionJustification{justified}{\leftskip=0pt \rightskip=0pt \parfillskip=0pt plus 1fil}
\definecolor{vividviolet}{rgb}{0.62, 0.0, 1.0}
\definecolor{amaranth}{rgb}{0.9, 0.17, 0.31}
\definecolor{palatinateblue}{rgb}{0.15, 0.23, 0.89}
\definecolor{brightpink}{rgb}{1.0, 0.0, 0.5}
\definecolor{cornflowerblue}{rgb}{0.39, 0.58, 0.93}
\definecolor{deepcarminepink}{rgb}{0.94, 0.19, 0.22}
\definecolor{radicalred}{rgb}{1.0, 0.21, 0.37}

\hypersetup{ linktoc=all,
    colorlinks, linkcolor={palatinateblue},
    citecolor={brightpink}, urlcolor={amaranth}
}

\graphicspath{{Images/}}

\newcommand{\red}[1]{\textcolor[rgb]{1.00,0.00,0.00}{#1}}

\graphicspath{{Images/}}

\def\@fnsymbol#1{\ensuremath{\ifcase#1\or \ddagger \or  $\textleaf$  \or \dagger
\else\@ctrerr\fi}}%

\makeatother

\def\sideremark#1{\ifvmode\leavevmode\fi\vadjust{\vbox to0pt{\vss
 \hbox to 0pt{\hskip\hsize\hskip1em
 \vbox{\hsize1.3cm\tiny\raggedright\pretolerance10000
 \noindent #1\hfill}\hss}\vbox to8pt{\vfil}\vss}}}%
\def\beq{\begin{equation}}
\def\eeq{\end{equation}}

 \newcommand{\be}{\begin{equation}}
	\newcommand{\en}{\end{equation}}

\begin{document}

\title{Phase Transitions and Critical Phenomena for the FRW Universe in an Effective \\ \vspace{0.2cm} Scalar-Tensor Theory}

\author{Haximjan Abdusattar}
\email{axim@nuaa.edu.cn}
\affiliation{College of Physics, Nanjing University of Aeronautics and Astronautics, Nanjing, 211106, China}
\affiliation{Key Laboratory of Aerospace Information Materials and Physics (NUAA), MIIT, Nanjing 211106, China}

\author{Shi-Bei Kong}
\email{kongshibei@nuaa.edu.cn}
\affiliation{College of Physics, Nanjing University of Aeronautics and Astronautics, Nanjing, 211106, China}
\affiliation{Key Laboratory of Aerospace Information Materials and Physics (NUAA), MIIT, Nanjing 211106, China}

\author{Hongsheng Zhang }
\email{sps\_zhanghs@ujn.edu.cn}
\affiliation{School of Physics and Technology, University of Jinan, 336 West Road of Nan Xinzhuang, Jinan, Shandong 250022, China}
\affiliation{Key Laboratory of Theoretical Physics, Institute of Theoretical Physics, Chinese Academy of Sciences, Beijing 100190, China}

\author{Ya-Peng Hu}
\email{huyp@nuaa.edu.cn}
\affiliation{College of Physics, Nanjing University of Aeronautics and Astronautics, Nanjing, 211106, China}
\affiliation{Key Laboratory of Aerospace Information Materials and Physics (NUAA), MIIT, Nanjing 211106, China}
\affiliation{Center for Gravitation and Cosmology, College of Physical Science and Technology, Yangzhou University, Yangzhou 225009, China}


\begin{abstract}

We find phase transitions and critical phenomena of the FRW (Friedmann-Robertson-Walker) universe in the framework of an effective scalar-tensor theory that belongs to the Horndeski class. We identify the thermodynamic pressure (generalized force) $P$ of the FRW universe in this theory with the work density $W$ of the perfect fluid, which is a natural definition directly read out from the first law of thermodynamics. We derive the thermodynamic equation of state $P=P(V, T)$ for the FRW universe in this theory and make a thorough discussion of its $P$-$V$ phase transitions and critical phenomena. We calculate the critical exponents, and show that they are the same with the mean field theory, and thus obey the scaling laws.


\end{abstract}

\maketitle

\section{Introduction}

Since 1970's, great efforts have been devoted to the investigations of black hole thermodynamics. Researchers come to several significant theoretical achievements in this area \cite{Hawking:1975vcx,Bekenstein:1974ax,Bardeen:1973gs}. Nowadays, black hole thermodynamics takes a fundamental status in different modern advances in theoretical physics, for example in AdS/CFT correspondence \cite{Maldacena:1997re,Witten:1998zw}. However, it is still unsatisfactory for the developments of black hole thermodynamics in aspect of observations. The principal reason roots in the global property of traditional black hole thermodynamics, which requires the knowledge of global structure of the manifold \cite{Gibbons:1977mu}. To apply black hole thermodynamics in realistic astrophysics, one needs to obtain the properties in future time-like infinity, null infinity, and space-like infinity \cite{Hayward:1994bu}. Besides gedanken experiments, it is impossible to obtain such information.

In view of this situation, researchers develop quasi-local black hole thermodynamics, which only needs the structure of a patch of the manifold. This development makes black hole thermodynamics become tractable in realistic astrophysical environments. A natural concept related to a patch of a manifold is apparent horizon \cite{Hayward:1994bu}. On the other hand, for black hole thermodynamics, the other crucial element is charge, and the desirable one is a conserved charge. Fortunately, for spherically symmetric spacetime, the Kodama vector leads to a conserved current, and further a conserved charge by integrating the current \cite{Jacobson:1995ab}. By using the properties of Kodama vector and apparent horizon, one arrives at the unified first law as a result of field equation \cite{Hayward:1993wb}.
In such an interpretation of field equation, a term related to a generalized force (thermodynamic pressure) appears in thermodynamic laws \cite{Padmanabhan:2003gd}.

FRW universe is a spherically symmetric spacetime, and has a thermal spectrum from its apparent horizon associated with a Hawking temperature \cite{Cai:2008gw,Hu:2010tx}. The existence of the apparent horizon is the main cause of having a self-consistent thermodynamics \cite{Gong:2007md}.
Ref.\cite{Cai:2005ra} first studied the connection between Friedmann equations and first law of thermodynamics for FRW universe \cite{Cai:2006rs}, and subsequently Refs.\cite{Akbar:2006kj,Akbar:2006mq} extended similar studies to some alternative theories of gravity.
However, some other important properties, including phase transition and critical behavior that have already been discovered in black holes, are rarely known for the FRW universe.

Note that, the thermodynamic pressure plays an important role for phase transitions and critical behaviors.
For example, in asymptotically AdS black hole, one usually treats the cosmological constant $\Lambda$ as the thermodynamic variable analogous to the pressure $P:=-{\Lambda}/{8\pi}$ \cite{Kastor:2009wy,Dolan:2010ha,Kubiznak:2012wp}, and its conjugate quantity is the thermodynamic volume $V$. Moreover, one can further construct an equation of state $P=P(V,T)$ for a black hole thermodynamic system, and investigate the $P$-$V$ (van der Waals-like \cite{DavidCJ:2014}) phase transition in the $P$-$V$ phase diagram \cite{Hu:2018qsy,Hu:2020pmr,Bhattacharya:2017hfj,Hendi:2012um}(see \cite{Gunasekaran:2012dq,Wei:2012ui,Cai:2013qga,Dehghani:2014caa,Xu:2015rfa,Spallucci:2013osa,Majhi:2016txt,Cheng:2016bpx,Dehyadegari:2017hvd,Estrada:2019cig,Frassino:2014pha,Hennigar:2016ekz,Dehghani:2022gwg,Hegde:2020xlv,Li:2020xkh}
for more related works, and also \cite{Altamirano:2014tva,Kubiznak:2016qmn} for reviews). For the FRW universe, the a significant issue is also related to the definition of its thermodynamic pressure $P$. In our recent papers \cite{Abdusattar:2021wfv,Kong:2021qiu,Kong:2022xny}, we have studied the first law of thermodynamics for the FRW universe and compared it with the usual standard form of first law $dU=TdS-PdV$, and hence identified the proper thermodynamic pressure $P$ with the work density $W$ of the matter field which defined by Hayward \cite{Hayward:1993wb}
\begin{equation}\label{W}
P\equiv W:=-h_{ab}T^{ab}/2\,,
\end{equation}
where $h_{ab}$ and $T^{ab}$ are the $0,1$-components of the metric and the stress-tensor \cite{Akbar:2006kj} with $a,b=0,1, x^0=t, x^1=r$.
\footnote{For asymptotically AdS black holes, the thermodynamic pressure can still be defined by the work density once
the cosmological constant term is treated as an effective stress-tensor, i.e. $T^{eff}_{\mu\nu}\equiv-\Lambda g_{\mu\nu}/8\pi$,
but the signs are different, i.e. $P\equiv-W$, which can be easily checked to be consistent with $P:=-{\Lambda}/{8\pi}$. It should be emphasized that this definition is more general, since it is independent from the existence of the cosmological constant~\cite{Abdusattar:2021wfv,Abdusattar:2022bpg}.}
Using this definition of thermodynamic pressure, we further derived the thermodynamic equation of state $P=P(V,T)$ for the FRW universe, and found an interesting $P$-$V$ phase transition in a gravity with a generalized conformal scalar field \cite{Kong:2021qiu}.
These results indicate that the FRW universe might has a similarity with usual van der Waals thermodynamic system, and it is interesting and worthy of further investigations of its phase transitions in other modified theories of gravity.


One of the most well-studied theories of modified gravity is the Lovelock gravity \cite{Lovelock:1971yv}, which is a natural generalization of Einstein's gravity. Because it gives covariant, conserved, second-order field equations, it is of particular interest. However, Lovelock terms usually do not have dynamical contribution to the field equations in four dimensional spacetime. Recently,
a trick has been proposed to circumvent this limitation \cite{Casalino:2020kbt,Casalino:2020pyv,Glavan:2019inb}. The trick is inspired by the novel 4D EGB gravity \cite{Glavan:2019inb}, where the coupling constant has been rescaled as $\alpha=\alpha'/(D-4)$, and taken the $D \rightarrow 4$ limit in the field equation. This procedure leaves nonvanishing contributions of the Lovelock terms on the equation of motion and thus one ends up with a seemingly novel theory of gravity in four dimension. However, this trick has been criticized to be ill-defined at the action level, and it can also result in divergence in the equations of motion and break the diffeomorphism of a general 4D spacetime \cite{Arrechea:2020gjw,Gurses:2020ofy,Hennigar:2020lsl}. To tackle this problem, a well-defined effective scalar tensor reformulation of Lovelock gravity was proposed \cite{Kobayashi:2020wqy}.

In the present paper, we would like to investigate the thermodynamics of the FRW universe in the effective scalar-tensor theory \cite{Kobayashi:2020wqy}. A scalar-tensor theory with high order derivatives is Horndeski gravity \cite{Kobayashi:2019hrl}, whose equations of motion only have second-order derivatives, which eliminates any Ostrogradsky instability \cite{Fernandes:2021dsb}, which is similar to Lovelock gravity \cite{Lovelock:1971yv}. In Horndeski gravity, the $P$-$V$ phase transition has been found to occur in black holes \cite{Hu:2018qsy}, which has raised interest in whether a similar phase transition can occur in the FRW universe. In this paper, we investigate the thermodynamic law of FRW universe in the most generic modified theory of gravity, i.e. effective scalar-tensor theory, and obtain its thermodynamic pressure $P$. Furthermore, we derive the thermodynamic equation of state $P=P(V,T)$ of FRW universe in this theory, and find that it also represents a phase transition and critical behaviour around the critical point.

This paper is organized as follows. In Sec.\ref{sec:Love}, we briefly review the effective scalar-tensor theory and its Friedmann equations in a FRW universe. In Sec.\ref{sec:PVT}, we investigate the thermodynamics of the FRW universe in this modified gravity, and derive its thermodynamic equation of state $P=P(V, T)$. In Sec.\ref{sec:Critical}, we demonstrate the $P$-$V$ phase transition and critical behaviors of the FRW universe in the effective scalar-tensor theory. In Sec.\ref{sec:con}, we make conclusions and discussion.

\section{A Brief Introduction \red{of the} Effective Scalar-Tensor Theory and FRW Universe}\label{sec:Love}

In this part we make a brief introduction on effective scalar-tensor reformulation of the regularized Lovelock gravity of Refs.\cite{Glavan:2019inb,Casalino:2020kbt,Casalino:2020pyv} and it's Friedmann's equation's in the FRW universe.

The action from which this theory\footnote{This theory can be viewed as a particular subclass of the Horndeski theory \cite{Gao:2020jhq,Horndeski:1974wa,Lu:2020iav,Fernandes:2020nbq} (See also \cite{Alkac:2022fuc,Aoki:2020lig} for more related works and for an extensive review \cite{Fernandes:2022zrq,Kobayashi:2019hrl} of the literature).} is given by \cite{Kobayashi:2020wqy}
\begin{equation}\label{action}
S = \int d^D x \,\sqrt{-g} \, \mathcal{L}\, + S_m\,,
\end{equation}
where $g$ is a determinant of the metric tensor $g_{\mu\nu}$, $S_m$ is the action associated with matter fields, and the Lagrangian
\footnote{If ${\alpha'}_3$ is set to zero, it is consistent with the result of the well-defined version of the four dimensional Einstein-Gauss-Bonnet theory \cite{Fernandes:2021dsb,Hennigar:2020lsl}.}
\begin{eqnarray}\label{STLL}
{\cal L}&=&\alpha_0+\left(\alpha_1-2 {\alpha'}_2 Ke^{-2\chi} \right){\mathcal R} +{\alpha'}_2 \left[-6K^2e^{-4\chi}+24Ke^{-2\chi}X+8X^2+8X\Box\chi +4G^{\mu\nu}\chi_\mu\chi_\nu
+\chi{\cal G}\right]\nonumber\\
&&+{\alpha'}_3\left[{\cal L}_{2}^H\{192X^3\}+{\cal L}_3^{H}\{-144 X^2\}+{\cal L}_4^H\{24X^2\}+{\cal L}_5^H\{48X\}\right]\,,
\end{eqnarray}
where $\mathcal R$ and $G_{\mu\nu}$ are the four-dimensional Ricci scalar and Einstein tensor, respectively, and $\cal G$ is the Gauss-Bonnet combination of the four-dimensional curvature tensors, ${\cal G}={\mathcal R}_{\mu \nu \gamma \delta}{\mathcal R}^{\mu \nu \gamma \delta}-4{\mathcal R}_{\mu \nu }{\mathcal R}^{\mu \nu}+{\mathcal R}^{2}$.
Here, $\chi$ is related to the metric in $n$-dimensional maximally symmetric space, $\chi_\mu\equiv\nabla_\mu\chi, X\equiv-\chi_\mu\chi^\mu/2$,
$K$ is the constant curvature, and ${\cal L}^H_i$ with $i=2,3,4,5$ are the corresponding terms in Horndeski theory. Note that the $\alpha_0$ and $\alpha_1$ correspond to Einstein's theory with the cosmological constant, ${\alpha'}_2$, ${\alpha'}_3$ are the coupling constants with dimension of $[length]^{2}$, $[length]^{4}$, and we denote them, as $\alpha$, $\beta$ for simplicity in the following calculation.

In the co-moving coordinate system $\{t,r,\theta,\varphi\}$, the line-element of FRW universe is written as
\begin{equation}\label{FRWds}
d s^2=-d t^2+a^2(t)\left[\frac{d r^2}{1-kr^2}+r^2(d\theta^2+\sin^2\theta d\varphi^2)\right]\,,
\end{equation}
where $a(t)$ is the time-dependent scale factor, $k$ is the spacial curvature.
The matter field in the FRW universe, is usually treated as a perfect fluid with stress-tensor
\begin{equation}\label{Tmunu}
T_{\mu\nu}=(\rho+p)u_{\mu}u_{\nu}+p g_{\mu\nu}\,,
\end{equation}
where $\rho$ and $p$ are the energy density and pressure, and $u_{\mu}$ is the four-velocity of the fluid satisfying $u_\mu u^\mu=-1$.

Applying the modified theory of gravity (\ref{action}) with (\ref{STLL}) to the spatially flat ($k=0$) (\ref{FRWds}) FRW universe with perfect fluid energy momentum stress-tensor (\ref{Tmunu}), the Friedmann's equations \cite{Kobayashi:2020wqy,Cai:2005ra,Gong:2007md} are obtained as
\begin{eqnarray}\label{FE10}
(1+\alpha H^2+\beta H^4)H^2&=&\frac{8\pi}{3}\rho \,,
\end{eqnarray}
and
\begin{eqnarray}\label{FE11}
(1+2\alpha H^2 +3\beta H^4)\dot{H}&=&-4\pi(\rho+p)\,,
\end{eqnarray}
which are also satisfy the energy conservation equation
\begin{eqnarray}\label{continutyEq}
\dot{\rho}+3H(\rho+p)=0\,,
\end{eqnarray}
where $H\equiv\dot{a}(t)/a(t)$ is the Hubble parameter, and $``\cdot"$ stands for the derivative with respect to the cosmic time.
It should be emphasized that, if $\beta=0$, the Eqs.(\ref{FE10}) and (\ref{FE11}) reduces to the Friedmann's equations obtained in holographic cosmology \cite{Apostolopoulos:2008ru,Bilic:2015uol,Lidsey:2009xz}, quantum corrected entropy-area relation \cite{Cai:2008ys}, four dimensional Einstein-Gauss-Bonnet gravity \cite{Feng:2020duo} and gravity with a generalized conformal scalar field \cite{Kong:2021qiu,Fernandes:2021dsb}.



\section{Thermodynamics and Equation of State for the FRW Universe in \red{the} Effective Scalar-Tensor Theory}\label{sec:PVT}

In this section, in the framework of the effective scalar-tensor theory, we give the first law of thermodynamics and construct an equation of state for the FRW universe from its Friedmann's equations.

For later convenience, we rewrite the line element of FRW universe (\ref{FRWds}) with areal radius $R\equiv a(t)r$ to the spatially flat ($k=0$) in the following form
\begin{equation}\label{NewMetric}
d s^2=h_{ab}d x^a d x^b+R^2(d\theta^2+\sin^2\theta d\varphi^2)\,,
\end{equation}
where $a,b=0,1$ with $x^0=t, x^1=r$ and $h_{ab}=[-1,a^2(t)]$. For simplicity, we denote $a \equiv a(t)$ in the following. For the dynamical spacetime there is an apparent horizon which the marginally trapped surface with vanishing expansion and satisfies $h^{ab}\partial_a R\partial_b R=0$ \cite{Hayward:1993wb}. Apply this condition to the metric (\ref{NewMetric}), one can easily obtain the radius of apparent horizon of the FRW universe \cite{Cai:2005ra}
\begin{equation}\label{AH}
R_A=\frac{1}{H}\,,
\end{equation}
whose time derivative is
\begin{equation}\label{dot}
\dot{R}_A=-H \dot{H} R^3_A \,,
\end{equation}
which characterizes the time-dependent nature of the apparent horizon.

With the expressions of the apparent horizon (\ref{AH}) and (\ref{dot}), one can rewrite the Friedmann's equations (\ref{FE10}) and (\ref{FE11}) as
\begin{eqnarray}
\rho&=&\frac{3}{8\pi R^2_A}\left(1+\frac{\alpha}{R^2_A}+\frac{\beta}{R^4_A}\right)\,,\label{rho1}\\
p&=&-\frac{3}{8\pi R^2_A}\left(1+\frac{\alpha}{R^2_A}+\frac{\beta}{R^4_A}\right)+\frac{\dot{R}_A}{4\pi HR^3_A} \left(1+\frac{2\alpha}{R^2_A}+\frac{3\beta}{R^4_A}\right)\,.\label{pm1}
\end{eqnarray}
Substituting the above expressions into Eq.(\ref{W}), we obtain the work density of the matter field for a FRW universe in this effective scalar-tensor theory
\begin{equation}\label{WD}
W=\frac{1}{2}(\rho-p)=\frac{3}{8\pi R^2_A}\left(1+\frac{\alpha}{R^2_A}+\frac{\beta}{R^4_A}\right)-\frac{\dot{R}_A}{8\pi HR^3_A} \left(1+\frac{2\alpha}{R^2_A}+\frac{3\beta}{R^4_A}\right) \,.
\end{equation}

The surface gravity on the apparent horizon of the spatially flat FRW universe \cite{Cai:2005ra}
\begin{equation}\label{sg}
\kappa=-\frac{1}{R_A}\Big(1-\frac{\dot{R}_A}{2}\Big)\,.
\end{equation}
Thus the Hawking temperature associated with the apparent horizon of the spatially flat FRW universe is \cite{Cai:2005ra}
\begin{equation}\label{HawkingT}
T\equiv\frac{|\kappa|}{2\pi}=\frac{1}{2\pi R_A}\Big(1-\frac{\dot{R}_A}{2}\Big)\,.
\end{equation}

The total energy of matter inside the apparent horizon is usually defined by $E=\rho V$ \cite{Gong:2007md,Akbar:2006kj}. Accordingly, we use the Eq.(\ref{rho1}) and thermodynamic volume $V={4\pi R_A^3}/{3}$ to obtain the energy for the FRW universe
\begin{equation}\label{EE}
E=\rho V=\frac{R_A}{2}+\frac{\alpha}{2R_A}+\frac{\beta}{2R^3_A}\,,
\end{equation}
which will actually lead to a good thermodynamic first law to the FRW universe in this scalar-tensor theory as
\footnote{This relation holds in nearly all of previous studies \cite{Abdusattar:2021wfv,Kong:2021qiu,Kong:2022xny}, where the energy $E$ could be regarded as an effective Misner-Sharp energy \cite{Maeda:2007uu,Cai:2009qf,Cai:2008mh}.}
\begin{eqnarray}\label{dEE}
dE=-TdS+WdV \,,
\end{eqnarray}
where $T$ is the Hawking temperature (\ref{HawkingT}), $W$ is the work density (\ref{WD}). One immediate result of the above relation is the explicit form of the entropy \begin{equation}\label{S}
S=\pi R_A^2+4\pi\alpha\ln{\frac{R_A}{R_0}}-\frac{3\pi\beta}{R_A^2}
=\frac{A}{4}+2\pi\alpha\ln\frac{A}{A_0}-\frac{12\pi^2\beta}{A}\,,
\end{equation}
where $A_0, R_0$ are constants.
The entropy $S$ includes three terms, the first term is the Bekenstein-Hawking entropy, the second term is a logarithmic correction which often appears as the leading-order quantum correction \cite{Cai:2009ua,Mukherji:2002de,Chatterjee:2003uv,Domagala:2004jt,Kaul:2000kf,Sen:2012dw,Ashtekar:1997yu,Rovelli:1996dv}, the third term represents further fluctuation of the entropy \cite{Sheykhi:2010wm,Zhu:2009qc,Fernandes:2020rpa} could be regarded as an effective theory of quantum gravity. Note that the Eq.(\ref{S}) recovers to the result of four dimensional regularized Gauss-Bonnet AdS black hole \cite{Fernandes:2021dsb,Fernandes:2020rpa}, if $\beta=0$.
It should also be pointed that the minus sign\footnote{The minus sign before $T$ is a common feature of cosmological horizons such as the apparent horizon of the FRW universe and the event horizon of the de Sitter spacetime, but its nature and interpretation is a longstanding and puzzling problem. Recently, some interesting papers \cite{Banihashemi:2022jys,Banihashemi:2022htw} were dedicated to clarify this problem.}
before $TdS$ in (\ref{dEE}) arises from the treatment that the surface gravity (\ref{sg}) on the apparent horizon is negative \cite{Abdusattar:2021wfv,Abdusattar:2022bpg,Dolan:2013ft}.

For a thermodynamic system, besides the laws of thermodynamics, equation of state like the van der Waals system usually also plays an important role. In order to clearly obtain the equation of state of the FRW universe in the effective scalar-tensor theory, we first compare Eq.(\ref{dEE}) with the standard form of the thermodynamic first law
\begin{equation}\label{SFL}
d U=Td S-Pd V \,,
\end{equation}
one can read out the internal energy $U$ and thermodynamic pressure $P$, i.e.
\begin{eqnarray} \label{PW}
U\equiv&-E\,, \\
P\equiv&W\,.
\end{eqnarray}
Using Eqs.(\ref{HawkingT}), (\ref{WD}) and (\ref{PW}), we further obtain the equation of state for the FRW universe in the effective scalar-tensor theory, i.e.\footnote{If $\beta$ is equal to zero, this equation reduces to that of a gravity with a generalized conformal scalar field \cite{Kong:2021qiu}, and if both $\alpha$ and $\beta$ are set to zero, it recovers the one in Einstein gravity \cite{Abdusattar:2021wfv}.}
\begin{eqnarray}\label{EoSLove}
P=\frac{T}{2 R_A}\left(1+\frac{2\alpha}{R^2_A}+\frac{3\beta}{R^4_A}\right)+\frac{1}{8\pi R^2_A}\left(1-\frac{\alpha}{R^2_A}-\frac{3\beta}{R^4_A}\right)\,,
\end{eqnarray}
where $R_A=(3V/4\pi)^{1/3}$.
The $\alpha$ and $\beta$ terms may contain some new features of the FRW universe,
which will be demonstrated in the following discussions. If $\alpha$ is absent, the equation of state is simplified to
\begin{equation}\label{EoSLovea0}
P=\frac{T}{2R_A}+\frac{1}{8\pi R^2_A}+\frac{3\beta T}{2R^5_A}-\frac{3\beta}{8\pi R^6_A}\,.
\end{equation}

\section{$P$-$V$ Criticality for \red{the} FRW Universe in the Effective Scalar-Tensor Theory}\label{sec:Critical}

In this section, we would like to study the $P$-$V$ phase transition and critical behaviors for the FRW universe in the effective scalar-tensor theory based on the equation of state (\ref{EoSLove}). We first obtain the critical point and illustrate the behaviors in the $P$-$V$ diagram. Then, we further calculate the critical exponents and discuss whether they satisfy the scaling laws or not.

\subsection{$P$-$V$ phase transition and critical behavior}

The necessary conditions for $P$-$V$ phase transition are \cite{Kubiznak:2012wp,Hu:2018qsy}
\begin{eqnarray}\label{PV}
\left(\frac{\partial P}{\partial V}\right)_{T}=\left(\frac{\partial^2 P}{\partial V^2}\right)_{T}=0\,,
\end{eqnarray}
or equivalently
\begin{eqnarray}\label{PVTc}
\left(\frac{\partial P}{\partial R_A}\right)_{T}=\left(\frac{\partial^2 P}{\partial R^2_A}\right)_{T}=0\,,
\end{eqnarray}
have a critical-point solution $T=T_c,\ P=P_c,\ R_A=R_c$.

For the equation of state (\ref{EoSLove}), the critical conditions (\ref{PVTc}) are
\begin{eqnarray}
\frac{9\beta}{4\pi R_c^7}-\frac{15\beta T_c}{2 R_c^6}+\frac{\alpha}{2\pi R_c^5}-\frac{3\alpha  T_c}{R_c^4}-\frac{1}{4\pi R_c^3}-\frac{T_c}{2 R_c^2}&=&0\,,\label{c1}
\\
-\frac{63\beta}{4\pi R_c^8}+\frac{45 \beta T_c}{R_c^7}-\frac{5\alpha}{2\pi R_c^6}+\frac{12\alpha  T_c}{R_c^5}+\frac{3}{4\pi R_c^4}+\frac{T_c}{R_c^3}&=&0\,.\label{c2}
\end{eqnarray}
From the above two equations (\ref{c1}) and (\ref{c2}), one can obtain the equation for the critical radius of the apparent horizon
\begin{equation}\label{RcEq}
-R_c^8+12\alpha R_c^6+12 \alpha^2 R_c^4+90 \beta R_c^4+132\alpha \beta R_c^2+135 \beta ^2=0\,.
\end{equation}
It can be seen that the Eq.(\ref{RcEq}) contains higher-order terms of $R_c$, so it is very difficult to solve analytically.
We first discuss a simple case, i.e. $\alpha=0, \beta\neq0$, and then the general case with $\alpha\neq0, \beta\neq0$.

%

\subsubsection{\large $\alpha=0, \quad \beta\neq0$}

In this situation, there is also a critical point if $\beta$ is negative:
\begin{eqnarray}\label{RcTcPc}
R_c=\sqrt[4]{3(15-4\sqrt{15})\beta}, \quad T_c=\frac{1}{2\times5^{3/4} \pi \sqrt[4]{(1-{4}/\sqrt{15})\beta}},
\quad P_c=-\frac{\sqrt{-(39+{152}/{\sqrt{15}})\beta}}{60\pi \beta},
\end{eqnarray}
where Eqs.(\ref{EoSLovea0}) and (\ref{PVTc}) have been used.
From the above results, one can still get a dimensionless constant:
\begin{equation}\label{RPTc}
\frac{2 R_{c}P_{c}}{T_{c}}=\frac{2}{3}+\frac{1}{\sqrt{15}}\approx 0.924866\,,
\end{equation}
which is much larger than $3/8$ (or 0.375) in the van der Waals system \cite{Kubiznak:2012wp,Bhattacharya:2017hfj}.

For convenience, one can define dimensionless pressure, temperature and radius as
\begin{equation}\label{reduce}
\widetilde{P}:=\frac{P}{P_{c}}\,, \quad \widetilde{R}:=\frac{R_A}{R_{c}}\,, \quad \widetilde{T}:=\frac{T}{T_{c}}\,.
\end{equation}
and rewrite (\ref{EoSLovea0}) as
\begin{equation}\label{EoSLovea01}
\widetilde{P}=\frac{2[3(4 \sqrt{5}-5 \sqrt{3}) \widetilde{R}^4-\sqrt{3}] \widetilde{R} \widetilde{T}+5 (4 \sqrt{3}-3 \sqrt{5}) \widetilde{R}^4+\sqrt{5}}{2 \left(5 \sqrt{5}-6 \sqrt{3}\right) \widetilde{R}^6}\,.
\end{equation}

\begin{figure}[h]
\begin{minipage}[t]{8cm}
\centering
\includegraphics[width=7cm,height =5.5cm]{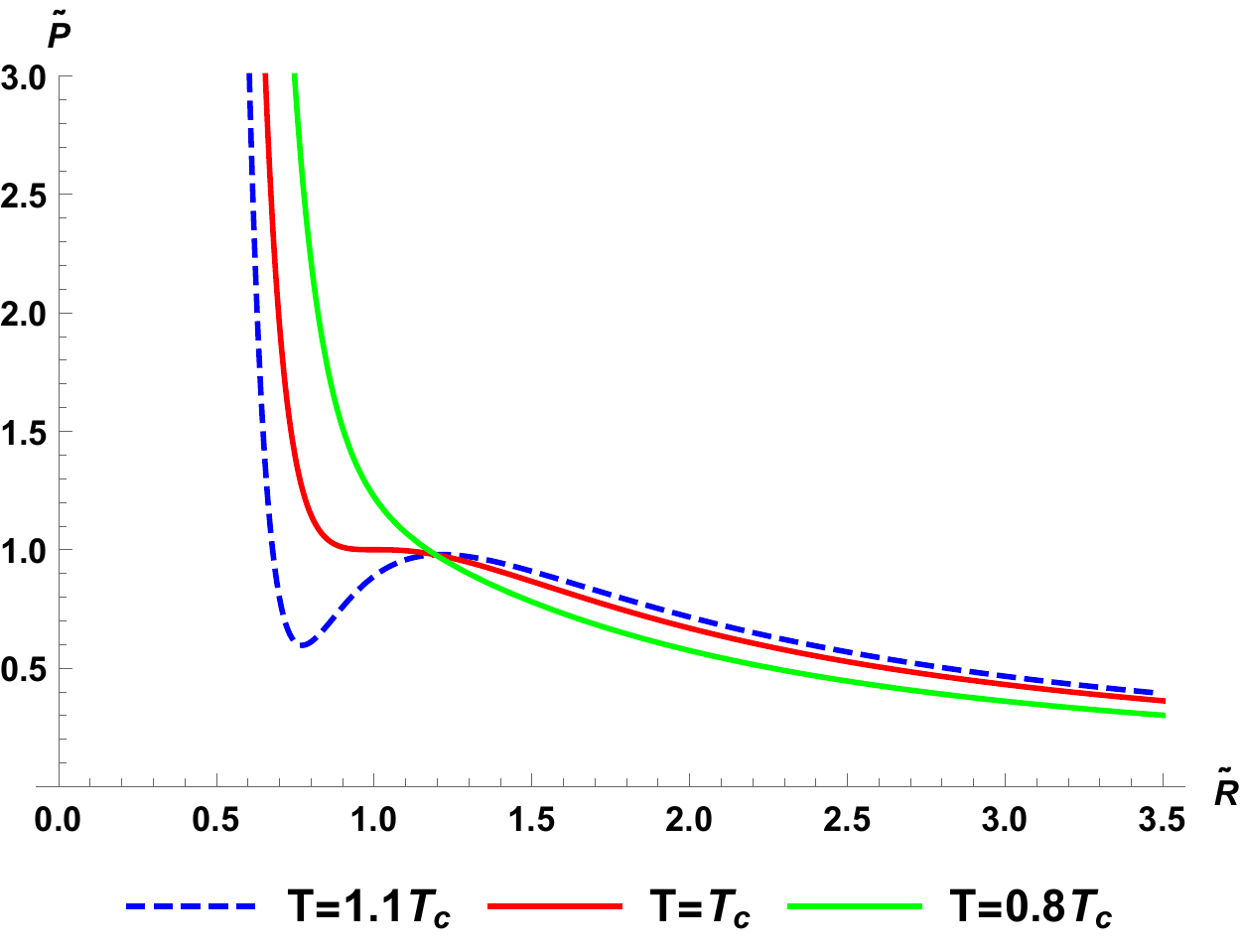}
\end{minipage}
\caption{Isothermal lines in the $\widetilde{P}$-$\widetilde{R}$ diagram. The dashed blue line is the isothermal line with a higher temperature,
i.e. $T > T_{c}$, where phase transition occurs; The red solid line is the isothermal line at the critical temperature $T = T_{c}$; The solid green line is the isothermal line with a lower temperature, i.e. $T < T_{c}$, where the system has only one phase thus no phase transition could occur. Note that the isothermal lines intersect at a thermodynamic singularity $\widetilde{R}_{s}=(1+4/15)^{1/4}$.}\label{Fig.EoSLovea01}
\end{figure}

In order to show the characteristic behaviors of the phase transition, we illustrate the corresponding $\widetilde{P}$-$\widetilde{R}$ diagram based on Eq.(\ref{EoSLovea01}) as shown in Fig.\ref{Fig.EoSLovea01}.

We can see from Fig.\ref{Fig.EoSLovea01} that, the phase transition occurs at the temperature larger\footnote{It means that the phase transition of the FRW universe is a high energy phenomenon, which probably happened at
the early stages of the Universe.} than the critical temperature $T > T_{c}$, while the behavior is similar to an ideal gas for $T < T_{c}$. This behavior is different from that of a van der Waals system and most of black holes system where coexistence phases appear below the critical temperature \cite{Kubiznak:2012wp,DavidCJ:2014,Hu:2018qsy,Bhattacharya:2017hfj}. Moreover, there is a `thermodynamic singularity', characterized by the pressure independence of temperature, i.e. $(\partial \widetilde{P}/\partial \widetilde{T})_{R_s}=0$. It leads to a common point for different isothermal lines in the $\widetilde{P}$-$\widetilde{R}$ diagram, which has also been found in many previous investigations  \cite{Frassino:2014pha,Hennigar:2016ekz,Li:2020xkh}.

\subsubsection{\large $\alpha\neq 0, \quad \beta\neq0$}

In this case, the analytical solution of the critical radius can hardly be acquired, but it is relatively easy to discuss the parameter space that allows a physical solution of the critical radius.

From Eqs.(\ref{EoSLove}), (\ref{PVTc}) and (\ref{RcEq}), one can get the formal solutions of the critical temperature and critical pressure:
\begin{eqnarray}\label{TcPc}
T_c=\frac{-R_c^4 +2\alpha R_c^2 +9\beta}{2\pi R_c(R_c^4+6\alpha R_c^2+15 \beta)},
\quad\quad\quad\quad P_c=-\frac{7 \alpha  R_c^6+2 R_c^4 (5 \alpha^2+33 \beta )+117\alpha \beta R_c^2 +126 \beta ^2}
{8\pi R_c^6 (R_c^4+6\alpha R_c^2+15\beta)}\,.
\end{eqnarray}
Note that the $R_c$, $T_c$ and $P_c$ are all should taken positive because of the critical point to be physical. This further constraints the allowed parameter space, see more detailed discussion in Appendix \ref{A}.
%
%
%
%
For $\alpha>0, \beta>0$, we can distinguish the critical pressure $P_c$ for given critical horizon radius $R_c$ is negative, which means the system has no physical critical point corresponding to a van der Waals like phase transition in this situation.
%
For $\alpha<0, \beta<0$, we find that when $\beta/\alpha^2\leq -\frac{8\sqrt{5}+15}{75}$, there is a critical point once the values of coupling constants $\alpha$ and $\beta$ are given. Therefore, we demonstrate some numerical results of these critical thermodynamic quantities as shown in Table \ref{TabRcTcPc}.

\renewcommand{\arraystretch}{1.6}
\begin{table}[!htbp]
\fontsize{10}{10}\selectfont
\caption{The critical thermodynamic quantities for some coupling constants $\alpha$ and $\beta$}
		\centering
\setlength{\tabcolsep}{3.3mm}{
		\begin{tabular}{p{1.2cm}<{\centering} p{1.4cm}<{\centering} p{1.8cm}<{\centering} p{1.8cm}<{\centering} p{1.8cm}<{\centering} p{1.8cm}<{\centering}}
			\hline
			\hline
			$   $   & $\beta$ & $R_{c}$ & $ T_c $ & $ P_c$ & $ R_c P_c/T_c$\\
   \hline
			\multirow{3}*{$\alpha=-1$}
            &$-2$~~~   & $1.551930$   & $0.075933$   & $0.019312$   & $0.394696$\\
            &$-1$~~~   & $1.388240$   & $0.083132$   & $0.022719$   & $0.373890$\\
&$-\frac{8\sqrt{5}+15}{75}$    & $1.242580$   & $0.089641$   & $0.026123$   & $0.362109$\\
			\hline
			\multirow{3}*{$\alpha=-\frac{1}{2}$}
            &$-2$~~~     & $1.430490$   & $0.084384$  & $0.024875$   & $0.421682$\\
            &$-1$~~~     & $1.245250$   & $0.096002$  & $0.031542$   & $0.409131$\\
&$-\frac{8\sqrt{5}+15}{300}$    & $0.878640$   & $0.126772$   & $0.052246$   & $0.362109$\\
			\hline\hline		
\end{tabular}
}
		\label{TabRcTcPc}
	\end{table}

We can see from Table \ref{TabRcTcPc} that for a specific $\alpha$ value, with the increase of $\beta$ value, the critical radius decreases, while the critical temperature and critical pressure increase, and the rates of them $R_{c}P_c/T_c$ decrease. In other words, when $\beta$ is chosen at a fixed value, we find that as $\alpha$ increases, the critical radius decreases, and the critical temperature and pressure, and the rates of them $R_{c}P_c/T_c$ are increase.

\subsection{critical exponents of the $P$-$V$ phase transition}

In this part, we calculate the critical exponents near the critical point of $P$-$V$ phase transition for a FRW universe in the framework of effective scalar-tensor theory.

For a thermodynamic system, near the critical point of phase transition, there are four critical exponents $(\widetilde{\alpha},\widetilde{\beta},\gamma,\delta)$ defined in the following \cite{Kubiznak:2012wp,Hu:2018qsy},
\begin{eqnarray}
C_{V}=T\left(\frac{\partial S}{\partial T}\right)_V\propto |\tau|^{-\widetilde{\alpha}}\,,~~~
\eta=\frac{V_l-V_s}{V_c} \sim {\omega_l -\omega_s} \propto |\tau|^{\widetilde{\beta}}\,,~~~
\kappa_T=-\frac{1}{V}\left(\frac{\partial V}{\partial P}\right)_T\propto |\tau|^{-\gamma}\,,~~~
\widetilde{P}-1\propto \omega^{\delta}\,,
\end{eqnarray}
with
\begin{eqnarray}
\tau=\frac{T}{T_c}-1\,,~~~~~~~~~~~~~~ \omega=\frac{R_A}{R_c}-1\,,
\end{eqnarray}
where the labels `s' and `l' stand for `small' and `large' respectively.

In the following, we will calculate the four critical exponents one by one for the FRW universe in the effective scalar-tensor theory.

As we have seen in previous section, the entropy of the FRW universe in the present work given in (\ref{S}) is only a function of the thermodynamic volume $V$ (or $R_A$). Therefore, one can know that the heat capacity at constant volume $C_V$ is zero, which suggests that the first critical exponent $\tilde{\alpha}=0$.
To obtain the other three critical exponents, we use the expand the equation of state (\ref{EoSLove}) near the critical point given by
\begin{eqnarray}\label{expansion}
\widetilde{P}=1+a_{10}\tau+a_{11}\tau\omega+a_{03}\omega^3+\mathcal{O}(\omega^4,\tau\omega^2)\,,
\end{eqnarray}
with coefficients
\begin{eqnarray}
a_{10}&=&\frac{T_{c}(R_{c}^4+2R_{c}^2\alpha+3\beta)}{2P_{c} R_{c}^5}\,,~~~~
a_{11}=-\frac{T_{c}(R_{c}^4+6R_{c}^2\alpha+15\beta)}{2P_{c} R_{c}^5}\,,~~~~
a_{03}=\frac{(R_{c}^4+2\alpha R_{c}^2 +3 \beta)(15\beta +\alpha R_{c}^2)}{\pi P_{c} R_{c}^6 (R_{c}^4+6 \alpha R_{c}^2+15\beta)}\,.\,\,\,~~~~
\end{eqnarray}
To clarity the critical behaviors of system, we demonstrate the coefficients of above equation in Table \ref{Taba11} by choosing some example values of the coupling constants ($\alpha$, $\beta$).
\renewcommand{\arraystretch}{1.5}
\begin{table}[!htbp]
\fontsize{10}{10}\selectfont
\caption{The coefficients in Eq.(\ref{expansion}) for some example values of $\alpha$ and $\beta$}
		\centering
\setlength{\tabcolsep}{3.3mm}{
		\begin{tabular}{p{1.2cm}<{\centering} p{1.4cm}<{\centering} p{1.4cm}<{\centering} p{1.4cm}<{\centering} p{1.4cm}<{\centering} p{1.4cm}<{\centering}}
			\hline
			\hline
			$   $   & $\beta$ & $a_{10}$ & $ a_{11} $ & $ a_{03}$ & $ a_{11}/a_{03}$\\
   \hline
			\multirow{3}*{$\alpha=0$}
            &$-2$~~~   & $-1.11672$   & $9.90859$   & $-6.39612$   & $-1.54915$\\
            &$-1$~~~   & $-1.11671$   & $9.90853$   & $-6.39580$   & $-1.54922$\\
			&$-0.5$    & $-1.11670$   & $9.90844$   & $-6.39565$   & $-1.54925$\\
			\hline
			\multirow{3}*{$\alpha=-1$}
            &$-2$~~~   & $-1.09543$   & $8.44041$   & $-4.96216$   & $-1.70084$\\
            &$-1$~~~   & $-1.11428$   & $8.10766$   & $-4.55391$   & $-1.78038$\\
&$-\frac{8\sqrt{5}+15}{75}$    & $-1.16977$   & $7.79486$   & $-4.03490$   & $-1.93186$\\
			\hline
			\multirow{3}*{$\alpha=-\frac{1}{2}$}
            &$-2$~~~     & $-1.09273$   & $9.04763$  & $-5.59590$    & $-1.61683$\\
            &$-1$~~~     & $-1.09078$   & $8.76603$  & $-5.31302$   & $-1.64991$\\
&$-\frac{8\sqrt{5}+15}{300}$    & $-1.16974$   & $7.79473$   & $-4.03461$   & $-1.93196$\\
			\hline\hline		
\end{tabular}
}
		\label{Taba11}
	\end{table}

According to the Maxwell's equal area law \cite{Spallucci:2013osa,Majhi:2016txt}, the end point of vapor and the starting point of liquid have the same pressure, i.e. $\widetilde{P}^*=\widetilde{P}_s=\widetilde{P}_l$, which indicates
\begin{eqnarray}
\widetilde{P}^*=a_{10}\tau+a_{11}\omega_{s}\tau+a_{03}\omega_{s}^3
=a_{10}\tau+a_{11}\tau\omega_{l}+a_{03}\omega_{l}^3\,,
\end{eqnarray}
or
\begin{eqnarray}\label{ea1}
a_{11}\tau(\omega_{l}-\omega_{s})+a_{03}(\omega_{l}-\omega_{s})^3=0\,.
\end{eqnarray}
Another relation from the Maxwell's equal area law is that $\int^s_l \widetilde{P}~dV=0$ in the $P$-$V$ phase diagram \cite{Hu:2018qsy,Spallucci:2013osa}, which gives the following equation
\begin{eqnarray}\label{ea2}
2a_{11}\tau(\omega_{l}^2-\omega_{s}^2)+3a_{03}(\omega_{l}^4-\omega_{s}^4)=0\,,
\end{eqnarray}
where Eq.(\ref{expansion}) has been used. From the above two Eqs. (\ref{ea1}) and (\ref{ea2}), one can get a nontrivial solution
\begin{eqnarray}
\omega_{l}=\sqrt{-\frac{a_{11}}{a_{03}}\tau}\,,\quad \quad \quad \omega_{s}=-\sqrt{-\frac{a_{11}}{a_{03}}\tau}\,,
\end{eqnarray}
so
\begin{eqnarray}
\omega_{l}-\omega_{s}=2\sqrt{-\frac{a_{11}}{a_{03}}\tau}\propto |\tau|^{1/2}\,,
\end{eqnarray}
which shows that the second critical exponent $\widetilde{\beta}=1/2$. Since $a_{11}/a_{03}$ is negative, we have $\tau>0$, which means that the coexistence phases in $P$-$V$ diagram appear above the critical temperature $T>T_{c}$.\footnote{This is different from the behavior of the
usual van der Waals systems and AdS black holes, where the coexistence phases are below the critical temperature \cite{Kubiznak:2012wp,DavidCJ:2014,Hu:2018qsy}.}

The isothermal compressibility near the critical point can be calculated as follows
\begin{eqnarray}
\kappa_T=-\left.\frac{1}{V_{c}}{\left(\frac{\partial V}{\partial \widetilde{P}}\right)_T}\right|_{c}\propto-\left(\frac{\partial \widetilde{P}}{\partial \omega}\right)_{\omega=0}^{-1}=-\frac{1}{a_{11}\tau}\propto \tau^{-1}\,,
\end{eqnarray}
which provides the third critical exponent $\gamma=1$.

When the temperature equals to the critical temperature $T=T_c$ or $\tau=0$, from (\ref{expansion}) we get
\begin{eqnarray}
\widetilde{P}-1\propto \omega^3\,,
\end{eqnarray}
which gives the fourth critical exponent $\delta=3$.

In summary, the four critical exponents associated with the $P$-$V$ phase transition of the FRW universe in the effective scalar-tensor theory are obtained by
\begin{eqnarray}
\widetilde{\alpha}=0\,,\quad\quad \widetilde{\beta}=\frac{1}{2}\,,\quad\quad \gamma=1\,, \quad\quad \delta=3\,,
\end{eqnarray}
which are consistent with the predications from the mean field theory \cite{Kubiznak:2012wp,DavidCJ:2014,Hu:2018qsy},
so the following scaling laws \footnote{It should be noted that there are only two independent ones.} are satisfied
\begin{eqnarray}\label{SL}
&&\widetilde{\alpha}+2\widetilde{\beta}+\gamma=2\,,~~~~~~~~\quad\quad\quad\quad\quad \widetilde{\alpha}+\widetilde{\beta}(1+\delta)=2\,,
\nonumber \\
&&\gamma(1+\delta)=(2-\widetilde{\alpha})(\delta-1)\,,~~~~~\quad\quad\quad \gamma=\widetilde{\beta}(\delta-1)\,.
\end{eqnarray}

\section{Conclusions and Discussion}\label{sec:con}

In this paper, we have studied the thermodynamic properties especially the $P$-$V$ phase transitions of the FRW universe with a perfect fluid in an interesting effective scalar-tensor theory. We have studied the first law of thermodynamic for the FRW universe in this theory, and identified the thermodynamic pressure $P$ with the work density of the perfect fluid, i.e.\ $P:=W$. Using this identification, we have further derived the thermodynamic equation of state for the FRW universe $P=P(V, T)$ in the effective scalar-tensor theory and found that there is a $P$-$V$ phase transition. However, unlike the van der Waals system and most of the black holes system, the phase transitions in this framework occur above the critical temperature. Last but not the least, we have also calculated the corresponding critical exponents and found that they are the same as those in the van der Waals system and mean field theory, so they satisfy the scaling laws.

Our results indicate that the van der Waals-type phase transition between a smaller phase and a larger phase for the FRW universe is a combined consequence of the matter fields, gravitational interactions and also the dynamical nature of the spacetime, etc.
It should be pointed out that the thermodynamic pressure of the FRW universe with perfect fluid is clearly compatible with the first law of thermodynamics and the Friedmann equations, and the construction of the equation of state also relies on the definitions of the thermodynamic variables $P, V, T$. We are also curious about when these phase transitions occur in the evolution of the real Universe, and whether we can detect them through cosmological observations. Our results provide a theoretical platform for future astronomical observations.
It would be also very interesting to study the thermodynamics and phase transitions of dynamical black holes in the same gravitational theory and compare the results with those of the FRW universe, and some universal properties may be found. These are open questions and will be studied in the future.

\section*{Acknowledgment}

We are grateful for the stimulating discussions with Profs. Li-Ming Cao, Theodore A. Jacobson, Xiao-Mei Kuang, Yen Chin Ong, Shao-Wen Wei. This work is supported by the National Natural Science Foundation of China (NSFC) under grants No.12175105, No.11575083, No.11565017.
H.Z. is Supported by the National Natural Science Foundation of China Grants No.12235019, and the National Key Research and Development Program of China (No. 2020YFC2201400).


\appendix

\section{Constraints on $\alpha$ and $\beta$ from Critical Point}\label{A}

In this appendix, based on Eqs.(\ref{RcEq}) and (\ref{TcPc}), we give conditions of the coupling constants $\alpha$ and $\beta$ for which the critical apparent horizon radius $R_c$, critical temperature $T_c$, and critical pressure $P_c$ are positive.

By introducing
\begin{equation}\label{Introduce}
 X=R_c^2>0\,,~~~~~~~~~~~ y=\frac{X}{\alpha}\,,~~~~~~~~~~~ \xi=\frac{\beta}{\alpha^2}\,,
\end{equation}
we can write Eq.(\ref{RcEq}) in the following form
\begin{equation}\label{RcEqXY}
-y^4+12y^3+12 y^2 +90 \xi y^2 +132\xi y +135 \xi^2=0\,,
\end{equation}
which has four roots $y_1, y_2, y_3, y_4$ that have complicated expressions, so we do not show them here. Since $X$ and $\alpha^2$ are positive, from Eq.(\ref{Introduce}) one can see that the signs of $y$ and $\xi$ are only determined by $\alpha$ and $\beta$ respectively. Based on this, we give the ranges of $\alpha$ and $\beta$ in the following way.

\begin{enumerate}
  \item If $\alpha\in(-\infty, 0)$ and $\beta/\alpha^2\in (-\infty, -(8\sqrt{5}+15)/75]\cup \beta/\alpha^2=(8\sqrt{5}-15)/75$,
  there is one positive critical radius $R_{c1}=\sqrt{\alpha y_1}$.

  \item If $\alpha\in(-\infty, 0)$ and $\beta/\alpha^2=(8\sqrt{5}-15)/75$, or
 if $\alpha\in(0,+\infty)$ and $\beta/\alpha^2\in (-\infty,-(8\sqrt{5}+15)/75]$,
  there is one positive critical radius $R_{c2}=\sqrt{\alpha y_2}$.

  \item If $\alpha\in(-\infty, 0)$ and $\beta/\alpha^2\in [(8\sqrt{5}-15)/75,1/15)$, or
  if $\alpha\in(0,+\infty)$ and $\beta/\alpha^2 =-(8\sqrt{5}+15)/75$,
  there is one positive critical radius $R_{c3}=\sqrt{\alpha y_3}$.

  \item If $\alpha\in(0, +\infty)$ and $\beta/\alpha^2=-(8\sqrt{5}+15)/75 \cup \beta/\alpha^2 \in [(8\sqrt{5}-15)/75,1/15)$,
  there is one positive critical radius $R_{c4}=\sqrt{\alpha y_4}$.
\end{enumerate}

The above results are shown in the following Table \ref{fuluTable}:
\begin{center}
\begin{tabular}{|c|c|c|c|c|c|c|c|c|}
\hline
conditions & $\xi<-\frac{8\sqrt{5}+15}{75}$ & $\xi=-\frac{8\sqrt{5}+15}{75}$  &$\xi=\frac{8\sqrt{5}-15}{75}$ & $\frac{8\sqrt{5}-15}{75}<\xi<\frac{1}{15}$ \\
\hline
$\alpha<0$ & $R_{c1}$ & $R_{c1}$  & $R_{c1},R_{c2},R_{c3}$  & $R_{c3}$ \\
\hline
$\alpha>0$ & $R_{c2}$ & $R_{c2},R_{c3},R_{c4}$  & $R_{c4}$ & $R_{c4}$  \\
\hline
\end{tabular}\label{fuluTable}
\end{center}

In our previous work \cite{Kong:2021qiu}, there is no $\beta$ term. In this case, the critical radius is very simple $R_c=\sqrt{(6-4\sqrt{3})\alpha}$. A natural question is that when $\beta$ vanishes, whether previous results can be recovered. Through careful analysis, we found that only $R_{c1}$ could revert to the previous result, which is the interesting result to us.
The root of (\ref{RcEqXY}) corresponding to $R_{c1}$ is $y_1$, which expression is

\begin{eqnarray}\label{y1}
 y_1=3-\frac{1}{\alpha^2}\sqrt{A\alpha^2+B}-\frac{1}{\alpha^2}\sqrt{2\alpha^2 A-B-\frac{24(3\alpha^2+7\beta)\alpha^4}{\sqrt{A\alpha^2+B}}}\,,
\end{eqnarray}
where
\begin{eqnarray}
A&\equiv&11\alpha^2+15\beta, \quad B\equiv\frac{\alpha^4-18\alpha^2\beta+45\beta^2}{\eta^{1/3}}+{\eta}^{1/3}\alpha^4 \,,
\nonumber \\
\eta&\equiv&\frac{9 \beta(3\alpha^4+17\alpha^2\beta-75 \beta^2)}{\alpha ^6}
-\frac{6}{\alpha^6}\sqrt{\beta ^2 (\alpha^2-3 \beta)^2(1075\beta^2+450\alpha^2\beta-19\alpha^4)}-1 \,. \nonumber
\end{eqnarray}

The critical temperature and critical pressure should be positive. When $\alpha>0, \beta>0$, we find from Eq.(\ref{TcPc}) that, the critical pressure is negative, i.e. $P_c<0$, so this case can be ruled out. Moreover, since we have chosen $R_{c1}$ as the critical horizon radius, we rule out the situation with $\alpha>0$.
For $\alpha<0, \beta>0$, from $T_c>0$ (in Eq.(\ref{TcPc})) we get
\begin{eqnarray}
  \Big(&&\alpha^2 > 3 \beta \cap \Big(R_c<\sqrt{-\sqrt{9 \alpha^2-15\beta}-3\alpha} \parallel \sqrt{\sqrt{\alpha^2+9 \beta}+\alpha}<R_c<\sqrt{\sqrt{9 \alpha^2-15 \beta}-3 \alpha}\Big)\Big)\parallel \nonumber\\
  \Big(&&\alpha^2 = 3 \beta \cap \Big(R_c <\sqrt{-\sqrt{9 \alpha^2-15 \beta}-3 \alpha}\parallel \sqrt{-\sqrt{9 \alpha ^2-15 \beta }-3 \alpha }<R_c<\sqrt{\sqrt{9 \alpha ^2-15 \beta }-3 \alpha }\Big)\Big) \parallel \nonumber\\
 \Big(&&\alpha^2 < 3 \beta \cap 3\alpha^2 > 5\beta \cap \Big(R_c<\sqrt{\sqrt{\alpha^2+9 \beta}+\alpha}\parallel \sqrt{-\sqrt{9 \alpha^2-15 \beta}-3 \alpha}<R_c<\sqrt{\sqrt{9 \alpha^2-15 \beta}-3 \alpha}\Big) \Big)\parallel ~~~~~~\nonumber\\
  \Big(&&3\alpha^2 \leq 5 \beta \cap \sqrt{\sqrt{\alpha^2+9 \beta }+\alpha }>R_c \Big)\,.
\end{eqnarray}
Unfortunately, it is difficult to obtain the constraints on $\alpha$ and $\beta$ from $P_c>0$. We think they are also determined by the critical temperature and pressure, so we don't consider the situation with $\beta>0$.

 For $\alpha<0, \beta<0$, from $T_c>0$ we get
 \begin{equation}\label{Tcdayu01}
 R_c<\sqrt{\sqrt{9 \alpha ^2-15 \beta }-3 \alpha }\,.
 \end{equation}
According to the Table \ref{fuluTable} and Eq.(\ref{Tcdayu01}) with the numerical analysis, we find that the critical temperature and pressure are positive, if $\xi=\beta/\alpha^2\leq -\frac{8\sqrt{5}+15}{75}$ in this situation.


\end{document}